\documentstyle{article}
\def\graphic #1#2#3#4#5{

    \noindent
    \centerline{\hrulefill}
    \leftline{\hbox to#1{\special{anisoscale #3, #1 #2}\hfil}}
    \vspace*{#2} \relax
    \vskip -3.9 cm
    \hskip 4.8 cm
    {\large \bf Universidade do Estado do Rio de Janeiro }
    \newline

    \vskip -0.25 cm
    \hskip 7.5 cm
    {\large \bf Instituto de F{\'\i}sica }

    \vskip 1 cm
    \hskip 7.5 cm
    {\large IF-UERJ-#4 }

    \hskip 7.5 cm
    {\large Preprint}

    \hskip 7.5 cm
    {\large #5 }

    \medskip
    \noindent
    \hrulefill

    \vskip 2.9 cm
    }
\def\({\c c}
\def\|{\'\i}

\def\dgraphic #1#2#3#4{
   \centerline{\hbox to#1{\special{anisoscale #3, #1 #2}\hfil}} 
   \vspace*{#2} \relax         
   \begin{figure}[h] \caption{#4} \end{figure}   
    }
\def\dtwographic #1#2#3#4{
    \centerline{\hbox to#1{\special{anisoscale #3, #1 #2}\hfil}
                \hbox to#1{\special{anisoscale #4, #1 #2}\hfil}}
    \vspace*{#2} \relax
    }

\def\dthreegraphic #1#2#3#4#5{
    \centerline{\hbox to#1{\special{anisoscale #3, #1 #2}\hfil}
                \hbox to#1{\special{anisoscale #4, #1 #2}\hfil}
                \hbox to#1{\special{anisoscale #5, #1 #2}\hfil}}
    \vspace*{#2} \relax
}

\def\ni{\noindent }
\def\eq #1{Eq.(\ref{#1})}
\def\l{\left}
\def\r{\right}

\def\st{$^{\rm st\,}$}

\def\nd{$^{\rm nd\,}$}
\def\ind{$^{\it \, nd\,}$}

\def\abaco#1{{\it abaco$_#1$}}
\def\bigcl{\left\{{\vrule height1.29em width0em depth1.29em}}
\def\bigcr{{\vrule height1.29em width0em depth1.29em}\right\}}
\def\bigql{\left[{\vrule height1.29em width0em depth1.29em}}
\def\bigqr{{\vrule height1.29em width0em depth1.29em}\right]}

\def\ni{\noindent }
\def\eq #1{Eq.(\ref{#1})}
\def\l{\left}
\def\r{\right}

\def\se #1{sec. \ref{#1}}
\def\st{$^{\rm st\,}$}

\def\nd{$^{\rm nd\,}$}
\def\ind{$^{\it \, nd\,}$}

\def\y1{\mbox{$y_{1}$}}
\def\yy2{\mbox{$y_{2}$}}

\def\ode{\mbox{\rm ode := \ \ }}
\def\X{\mbox{\rm X := \ \ }}
\def\abaco#1{{\it abaco$_#1$}}
\def\bigcl{\left\{{\vrule height1.29em width0em depth1.29em}}
\def\bigcr{{\vrule height1.29em width0em depth1.29em}\right\}}
\def\bigql{\left[{\vrule height1.29em width0em depth1.29em}}
\def\bigqr{{\vrule height1.29em width0em depth1.29em}\right]}

\begin{document}
\hspace\parindent
\thispagestyle{empty}

\graphic{2 in}{1.6 in}{uerj.wmf}{1/97}{March 1997}
\centerline{\LARGE \bf Computer Algebra Solving of Second Order ODEs}

\bigskip
\centerline{\LARGE \bf Using Symmetry Methods}

\bigskip
\bigskip
\centerline{\large
E.S. Cheb-Terrab\footnote{Symbolic Computation Group, Departamento de F\|sica
Te\'orica, IF-UERJ.
\newline
\hspace*{.55cm}Available as http://dft.if.uerj.br/preprint/e7-1.tex},
{L.G.S. Duarte}\footnotemark[2] and {L.A.C.P. da Mota}\footnotemark[1]}

\bigskip
\bigskip
\bigskip
\begin{large}
\begin{abstract}

An update of the {\it ODEtools} Maple package, for the analytical solving
of 1\st and 2\nd order ODEs using Lie group symmetry methods, is
presented. The set of routines includes an ODE-solver and user-level
commands realizing most of the relevant steps of the symmetry scheme. The
package also includes commands for testing the returned results, and for
classifying 1\st and 2\nd order ODEs.

\end{abstract}
\bigskip
\centerline{ \underline{\hspace{6.5 cm}} }

\medskip
\centerline{ 
{\bf (Accepted for publication in Computer Physics Communications)} }

\end{large}
\newpage
\setcounter{page}{1}

\bigskip
\bigskip
\hspace{1pc}
{\bf PROGRAM SUMMARY}
\bigskip

\begin{small}
\noindent
{\em Title of the software package:}  ODEtools.   \\[10pt]
{\em Catalogue number:} (supplied by Elsevier)                \\[10pt]
{\em Software obtainable from:} CPC Program Library, Queen's
University of Belfast, N. Ireland (see application form in this issue)
\\[10pt]
{\em Licensing provisions:} none  \\[10pt]
{\em Operating systems under which the program has been tested:}
UNIX systems, Macintosh, DOS (AT 386, 486 and Pentium based) systems,
DEC VMS, IBM CMS.                        \\[10pt]
{\em Programming language used:} {\bf Maple V} Release 3/4. \\[10pt]
{\em Memory required to execute with typical data:}  16 Megabytes. \\[10pt]
{\em No. of lines in distributed program, excluding On-Line Help,
etc.:} 10159. \\[10pt]
{\em Nature of mathematical problem}\\
Analytical solving of 1\st and 2\nd order ordinary differential equations
using symmetry methods, and the inverse problem; that is: given a set of 
point
and/or dynamical symmetries, to find the most general invariant 1\st or 2\nd
order ODE.\\[10pt]
{\em Methods of solution}\\
Computer algebra implementation of Lie group symmetry methods.   \\[10pt]
{\em Restrictions concerning the complexity of the problem}\\
Besides the inherent restrictions of the method (there is as yet no general
scheme for solving the associated PDE for the coefficients of the
infinitesimal symmetry generator), the present implementation does not work
with systems of ODEs nor with ODEs of differential order higher than two. 
\\[10pt]
{\em Typical running time}\\
This depends strongly on the ODE to be solved. For the case of first order
ODEs, it usually takes from a few seconds to 1 or 2 minutes. In the tests we
ran with the first 500 1\st order ODEs from Kamke's book \cite{kamke}, the
{\it average times} were: 8 sec. for a solved ODE and 15 sec. for an
unsolved one, using a Pentium 200 with 64 Mb. RAM, on a {\it Windows 95}
platform. In the case of second order ODEs, the average times for the
non-linear 2\nd order examples of Kamke's Book were 35 seconds for a solved
ODE and 50 seconds for an unsolved one. The tests were run using the Maple
version under development, but almost equivalent results are obtained using
the available Maple R4 and R3 (the code presented in this work runs in all
these versions). \\[10pt]
{\em Unusual features of the program}\\
The ODE-solver here presented is an implementation of all the steps of the
symmetry method solving scheme; that is, the command receives an ODE, and
when successful it directly returns a closed form solution for the
undetermined function. Also, this solver permits the user to {\it
optionally} participate in the solving process by giving advice concerning
the {\it functional form} for the coefficients of the infinitesimal symmetry
generator (infinitesimals). Many of the intermediate steps of the symmetry
scheme are available as user-level commands too. Using the package's
commands, it is then possible to obtain the infinitesimals, the related
canonical coordinates, the finite form of the related group transformation
equations, etc. Routines for testing the returned results, especially when
they come in implicit form, are also provided. Special efforts were put in
commands for solving the inverse problem too; that is, commands returning
the most general 1\st or 2\nd order ODE simultaneously invariant under given
symmetries. One of the striking new features of the package -related to 2\nd
order ODEs- is its ability to deal with dynamical symmetries, both in
finding them and in using them in the integration procedures. Finally, the
package also includes a command for classifying ODEs, optionally popping up
Help pages based on Kamke's advice for solving them, facilitating the study
of a given ODE and the use of the package with pedagogical purposes.

\end{small}
\newpage
\setcounter{page}{1}
%
%
\hspace{1pc}
{\large {\bf LONG WRITE-UP}}

\section*{Introduction}

In a previous work \cite{odetools1}, we presented an implementation of Lie
symmetry methods (SM) for solving first order ODEs. The key idea of that
work was to prepare routines for finding particular solutions for the PDE
determining the coefficients of the infinitesimal symmetry generators
(infinitesimals), as well as providing extra routines for the user's input
of {\it functional form} ans\"atze when the default routines fail. This
approach is presently solving 85\% of Kamke's examples using only the {\it
defaults}, apart from being a concrete way to tackle {\it
non-classifiable} 1\st order ODEs, for which the standard computer algebra
solvers usually fail.

The same idea can be implemented for 2\nd order ODEs too. To understand
the motivation, we recall that most implementations of SM for high order
ODEs are based on the setup and solving of the so-called {\it determining
equations} for the infinitesimals -an overdetermined system of PDEs-,
which arise when we assume we are interested only in point
symmetries \cite{hereman}. However, this poses a limitation on the 
2\nd order ODEs that can be dealt with, since only a restricted subset of
them have point symmetries. On the other hand, all ODEs have
infinite dynamical symmetries \cite{gon,olver2}, which arise as particular
solutions for a single linear 2\nd order PDE for the infinitesimals.

This paper then presents the implementation of the ideas of our previous
work to tackle 2\nd order ODEs. One of the tricky things related to the use
of SM to solve {\it n$^{th}$} order ODEs is that the knowledge of {\it n}
symmetries does not directly reduce the problem to a line integral as in the
1\st order case. Moreover, the alternatives found in the literature for
constructing the solution departing from dynamical symmetries are few. We
then extended some of the standard integration methods for point symmetries,
and implemented them as routines for dynamical symmetries too.

As a second issue, we invested in the {\it research} design of the package,
extending both the number and the capabilities of extra user-lever routines
related to the intermediate steps of the symmetry scheme. Worth mentioning
are a routine for finding the most general 1\st/2\nd order ODE
simultaneously invariant under many point or dynamical symmetries, and a
routine returning the symmetries of an {\it unknown} ODE, given its
solution.

As a third issue, we invested in augmenting the pedagogical potential of the
package by extending the classification capabilities of the {\bf odeadvisor}
command to work with most of the standard classifications for 2\nd order
ODEs (see \cite{zwillinger}).

The exposition is organized as follows. In \se{liemethod}, the SM scheme for
solving 2\nd order ODEs is briefly reviewed. In \se{package}, a compact
table-summary of all {\it ODEtools} routines and a detailed description of
the ODE-solver are presented. Sec. \ref{algorithms} briefly illustrates the
extension of the methods presented in \cite{odetools1} for finding the
infinitesimals. In \se{integrating}, we comment about the methods
implemented for integrating ODEs from the knowledge of their symmetries,
focusing on the case in which these symmetries are of dynamical type. Sec.
\ref{tests} displays the results of testing the package with the non-linear
2\nd order examples of Kamke's book as well as an update of the results
obtained for the Kamke's first 500 1\st order ODEs. In \se{compare}, the
main differences between {\it ODEtools} and other existing packages for
symmetry analysis of ODEs are highlighted. Finally, the conclusions contain
a brief discussion about this work and its possible extensions.

\section{Symmetry methods for 2\nd order ODEs}
\label{liemethod}


Generally speaking, the key point of Lie's method for solving ODEs is that
the knowledge of a (Lie) group of transformations which leaves a given ODE
invariant may help in reducing the problem of finding its solution to
quadratures \cite{stephani,bluman,olver}. Aside from the subtleties which
arise when considering different cases, we can summarize the computational
task of using SM for solving a given 2\nd order ODE, say,

\begin{equation}
{\frac {d ^{2}y}{d x^{2}}}\,=
\Phi (x, \,y, \,{\frac {d y}{d x}})
\label{ode}
\end{equation}

\ni as the finding of the infinitesimals of a one-parameter Lie group
which leaves \eq{ode} invariant, i.e., a pair of functions\footnote{In
what follows, $\y1=\frac{dy}{dx}$.} $\{\xi(y,x,\y1),\,\eta(y,x,\y1)\}$
satisfying

\begin{equation}
\begin{array}{l}
\displaystyle
\l(2\,\y1\, {\frac {\partial ^{2}\,\eta }{\partial y \,\partial \y1}} + 2\,
{\frac {\partial ^{2}\,\eta }{ \partial x\,\partial \y1}} - 3\,\y1\,
{\frac {\partial \,\xi}{\partial y}}  - 2\,{\frac {\partial \,\xi}{\partial
x}} - 2\,\y1\, {\frac {\partial^{2}\,\xi }{ \partial x\,\partial \y1}}
+ {\frac {\partial \,\eta }{\partial y}} - 2\, \y1^{2}\, {\frac
{\partial ^{2}\,\xi }{\partial y\,\partial \y1}} \r)\Phi \mbox{}
\\*[.25in]
\displaystyle
+ \l(
- \y1\, {\frac {\partial \, \eta }{\partial y}} + \y1^{2}\, {\frac {\partial
\,\xi }{\partial y}} + \y1\, {\frac {\partial \,\xi }{\partial x}} - {\frac
{\partial \,\eta }{\partial x}}\r)\, {\frac {\partial \,\Phi }{ \partial
\y1}} + {\frac {\partial ^{2}\,\eta }{\partial x^{2}}} + {\frac {\partial
^{2}\,\eta }{\partial y^{2}}} \,\y1^{2}
\\*[.25in]
\displaystyle
+ \l(
\y1 {\frac {\partial \,\eta }{\partial \y1}}
- \y1^{2} {\frac {\partial \,\xi }{\partial \y1}}
- \eta \r)
{\frac{\partial \,\Phi }{\partial y}}
+ \l( {\frac {\partial^{2}\,\eta }{\partial \y1^{2}}}
- \y1 {\frac {\partial ^{2}\, \xi }{\partial \y1^{2}}}
- 2{\frac {\partial \,\xi }{\partial \y1}} \r) \Phi ^{2}
\\*[.25in]
\displaystyle
+\l( {\frac
{\partial \,\eta }{\partial \y1}} - \y1 {\frac {\partial \,\xi }{\partial
\y1}} - \xi \r) {\frac {\partial \, \Phi }{\partial x}}
- {\frac {\partial
^{2}\,\xi }{\partial y^{2}}} \, \y1^{3} - 2\, {\frac {\partial^{2}\,\xi
}{\partial y\,\partial x}} \,\y1^{2} + 2\, {\frac {\partial^{2}\,\eta
}{\partial y\,\partial x}} \,\y1 - \y1\, {\frac { \partial^{2}\,\xi
}{\partial x^{2}}} =0
\end{array}
\label{xi_eta}
\end{equation}
followed by the integration of the ODE by either: 

\begin{description} 

\item [a)] determining the differential invariants of order 0 and 1 of the
symmetry group;

\item [b)] determining first integrals $\psi(x,y,\y1)=\psi_0$ (from the
knowledge of $\xi$ and $\eta$);

\item [c)] determining the canonical coordinates, say $\{r,s(r)\}$, of the
associated Lie group.

\end{description}


A first look at the symmetry scheme may lead to the conclusion that
finding solutions to \eq{xi_eta} would be much more difficult than solving
the original \eq{ode}. However, what we are really looking for is a
particular solution to \eq{xi_eta}, and in many cases this particular
solution is the only thing one can actually obtain. As an example,
consider

\begin{equation} \ode {\frac {d^{2}y}{d{x}^{2}}}= {\frac {\left (
\displaystyle {\frac {dy}{dx}}x-y\right )^{2}}{{x}^{3}}} \label{ode_1}
\end{equation} This 2\nd order ODE is non-linear and it doesn't match any
pattern for which we know the solving method {\it a priori}; standard
classification based ODE-solvers then fail when trying to solve it.
However, a polynomial {\it ansatz} for the infinitesimals (here made by
{\bf symgen}, the routine for determining the symmetries) rapidly leads to
3 particular solutions to \eq{xi_eta}\footnote{In what follows, the {\it
input} can be recognized by the Maple prompt \verb->-.}:

\begin{verbatim}
> symgen(ode);        # input = ODE,  output = pairs of infinitesimals
\end{verbatim}

\begin{equation}
[\xi=0,\eta=x],[\xi=x,\eta=y],[\xi={x}^{2},\eta=yx]
\label{sym1}
\end{equation}
Passing the ODE directly to {\bf odsolve} (the solver), it will internally
call {\bf symgen} and use the result above to solve the ODE as follows:

\begin{verbatim}
> odsolve(ode);
\end{verbatim}
\begin{equation}
y=\left (\ln (x) - \ln (1+xC_{1}) + C_{2} \right )x
\label{ans_1}
\end{equation}
What is amazing, and characteristic of symmetry methods is that if we change
$\l( {\frac {dy}{dx}}x-y\r)/x$ to $F\l(\l( {\frac {dy}{dx}}x-y\r)/x\r)$ in
\eq{ode_1}, where $F$ is an arbitrary function of its argument, the first
two symmetries of \eq{sym1} will remain valid and the solving scheme will
succeed as well:

\begin{equation}
\ode
{\frac {d^{2}y}{d{x}^{2}}}=
\frac{1}{x}\,\,
F\l(
\left (
\displaystyle
{\frac {dy}{dx}}\right )-
{\frac
{y}{x}}\r) 
\label{ode_2}
\end{equation}
\begin{verbatim}
> odsolve(ode);
\end{verbatim}
\begin{equation}
y=\left (
\displaystyle
\int
{
\displaystyle
{\rm RootOf}\l(
\ln (x)+C_{1}
+{\int}^{{\it \_Z}}
\frac{1}{\l({\it \_a}-F({\it \_a})\r)}\,{d\_a}
\r)}
\frac{dx}{x}
+C_{2}\right )x
\label{ans_2}
\end{equation}
The answer above is expressed using {\bf RootOf}, and the inner integral
uses the new {\bf intat} command\footnote{ {\bf intat} is a command of the
last version of {\it PDEtools} \cite{pdetools}, and represents an {\it
integral evaluated at a point} -analogous to a derivative evaluated at a
point. {\bf intat} displays the evaluation point as an upper limit of
integration.}. This kind of general answer can be interpreted as a {\it
mapping} in that, given $F$, it returns the answer after calculating the
roots of the resulting expression; for example, the ODE \eq{ode_1} is a
particular case of \eq{ode_2}, and its answer \eq{ans_1} is what one would
obtain introducing $F: u \rightarrow u^2$ in \eq{ans_2}.

Also, due to the fact that we are just looking for particular solutions to
\eq{xi_eta}, symmetry methods can be an alternative in solving linear ODEs
too. Consider, for instance:
\begin{equation}
\ode
{\frac {d^{2}y}{d{x}^{2}}}=
F(x){\l(\l({\frac {dy}{dx}}\r)-\frac{y}{x}\r)}
\label{linear_ODE}
\end{equation}
This ODE does not match a known pattern related to special functions, nor
can it be solved using the standard schemes for rational or exponential
solutions, due to the presence of an arbitrary function $F(x)$. However, its
invariance under $[\xi=0,\,\eta=x]$ and $[\xi=0,\,\eta=y]$ can be easily
determined by a polynomial {\it ansatz}, from which its solution follows
straightforwardly:
\begin{verbatim}
> odsolve(ode);
\end{verbatim}
\begin{equation}
y=x\l(C_{1}\,\int \!
{e^{^{\l(
\displaystyle
\int \!{\frac {x\,F(x)-2}{x}}{dx}\r)}}}{dx}+C_{2}\r)
\label{ans_linear_ODE}
\end{equation}

\section{The package's commands}
\label{package}

A detailed description of the package's commands, with examples and
explanations concerning their calling sequences, is found in the On-Line
Help, and is already present in \cite{odetools1}. Therefore, we have
restricted this section to a brief table-summary and a detailed description
only of the solver, {\bf odsolve}. Some {\it input/output} examples can be
seen in \se{algorithms} and \ref{integrating}.

\subsection{\it Summary}
A compact summary of the commands of the package is as follows:

\bigskip
{\centerline
{\small
\begin{tabular}{|p{1.2 in} |l|}
\hline
Command 	  & Purpose:\\
\hline
{\bf odsolve} & solves ODEs using the symmetry method scheme \\
{\bf intfactor } & looks for an integrating factor for first order ODEs\\
{\bf canoni } & looks for a pair of canonical coordinates of a given 
Lie group\\
{\bf eta\_k}   & returns the k-extended infinitesimal\\
{\bf transinv}  & looks for the finite group transformation\\
{\bf infgen}  & returns the k-extended symmetry generator as an operator\\
{\bf symgen}  & looks for pairs of infinitesimals\\
{\bf equinv}  & looks for the most general ODE invariant under a given
set of symmetries \\
{\bf buildsym } & looks for the infinitesimals given the solution of an 
ODE\\
{\bf odepde}  & returns the PDE for the infinitesimals\\
{\bf odetest} & tests explicit/implicit results obtained by 
ODE-solvers \\
{\bf symtest} & tests a given symmetry w.r.t a given ODE\\
{\bf odeadvisor}  & classifies 1\st/2\nd order ODEs and pop up
                related Help-pages\\
\hline
\multicolumn{2}{c}{Table 1. \it Summary of the {\it ODEtools} commands}\\
\end{tabular} }}

\subsection{\it Description}
\label{description}

\ni {\it Command name: {\bf odsolve}}
\label{odsolve}

\ni {\it Feature:} 1\st and 2\nd order ODE-solver based on symmetry methods

\ni {\it Calling sequence:}
\begin{verbatim}
> odsolve(ode);
> odsolve(ode, y(x), way=xxx, HINT=[[expr1, expr2],..], int_scheme);
\end{verbatim}

{\it Parameters:}

\noindent
\begin{tabular}{p{2.22in}p{3.8in}}
\verb-ode- & - a 1\st or 2\nd order ODE.\\

\verb-y(x)- & - the dependent variable (required when not obvious).\\

\verb-way=xxx- & - optional, forces the use of only one ({\tt xxx}) of the 8
internal algorithms \{abaco1, 2, 3, 4, 5, 6, abaco2, pdsolve\} for
determining the coefficients of the infinitesimal symmetry generator
(infinitesimals).\\

\verb-HINT = [e1,e2] - & - optional, {\tt e1} and  {\tt e2}, indicate 
possible {\it functional} forms for the infinitesimals.\\

\verb-HINT=[[e1,e2],[e3,e4],..] - & - optional, a list of {\it hints} for
the infinitesimals.\\

\verb-int_scheme-& - optional, one of: fat, can, can2, gon, gon2,
dif.
\end{tabular}

\noindent
Optional parameters can be given alone or in conjunction, and in any order.

\bigskip
\noindent
{\it Synopsis:}

Given a 1\st or 2\nd order ODE, {\bf odsolve}'s main goal is to solve it
in two steps: first, determine pairs of infinitesimals of 1-parameter
symmetry groups which leave the ODE invariant, and then use these
infinitesimals to integrate the ODE.

To determine the infinitesimals, {\bf odsolve} makes calls to {\bf symgen},
another command of the package. To integrate the ODE using these
infinitesimals, {\bf odsolve} has seven schemes, almost all of them
explained in connection with point symmetries in \cite{stephani,bluman}:

\begin{enumerate}

\item building an integrating factor (\verb-fat-, only for 1st order ODEs)
\label{int_fat}

\item reducing the ODE to a quadrature using the canonical coordinates of
that group (\verb-can-)
\label{int_can}

\item reducing a 2\nd order ODE to a quadrature at once, using 2 pairs of
infinitesimals forming a 2-D subalgebra (\verb-can2-)
\label{int_can2}

\item reducing a 2\nd order ODE to a quadrature, constructively, using a
normal form of the generator in the space of first integrals (\verb-gon-)
\label{int_gon}

\item reducing a 2\nd order ODE to a quadrature at once, using 2 pairs of
infinitesimals (a 2-D subalgebra) and normal forms of generators in the
space of first integrals (\verb-gon2-)
\label{int_gon2}

\item using differential invariants constructively (\verb-dif-).
\label{int_dif}

\item solving a 2nd order ODE at once, using 3 pairs of infinitesimals, when
no two of them can be used to form a 2-D subalgebra. \label{int_can3}

\end{enumerate}

The integration schemes (\ref{int_fat}) and (\ref{int_can}) are used
with 1\st order ODEs, while schemes (\ref{int_can}) to (\ref{int_can3}) work
with 2\nd order ODEs. The integration schemes (\ref{int_gon}),
(\ref{int_gon2}) and (\ref{int_dif}) work with dynamical symmetries too.
{\bf odsolve} does not classify the ODE before tackling it and is mainly
concerned with non-classifiable ODEs for which the standard Maple {\bf
dsolve} fails. By default, {\bf odsolve} starts off trying to isolate the
derivative in the given ODE, then sequentially uses subsets of the
algorithms of {\bf symgen} to try to determine the infinitesimals, and
finally sequentially tries the integration schemes mentioned above. The
default order for trying these schemes is

\begin{itemize}

\item 1\st order ODEs: can, fat

\item 2\nd order ODEs: gon, can2, gon2, can, dif

\end{itemize}

\ni When {\bf odsolve} succeeds in solving the ODE, it returns, in order of
preference:

\begin{itemize}

\item an explicit closed form solution;

\item an implicit closed form solution;

\item a parametric solution (default strategy for 1\st order ODEs when the
derivative cannot be isolated).

\end{itemize}

\ni All these defaults can be changed by the user; the main options 
she/he has are:

\begin{itemize}

\item To request the use of only one of the algorithms for determining the
infinitesimals (way=xxx option; different algorithms may lead to different
symmetries for one and the same ODE, sometimes making the integration step
easier).

\item To enforce the use, in a specific order, of only one or more of the
alternative schemes for integrating a given ODE after finding the
infinitesimals (fat, can, can2, gon, gon2, and dif optional arguments,
useful to select the best integration strategy for each case).

\item To indicate a possible functional form for the infinitesimals 
(HINT=xxx option). This option is valuable when the solver fails, or to 
study the possible connection between the algebraic pattern of the 
given ODE and that of the symmetry generators.

\end{itemize}


A brief description of how the \verb-HINT=xxx- option can be used is as 
follows:

\begin{itemize}

\item \verb-HINT=[e1,e2]-, indicates to the solver that it should take e1 
and e2 as the infinitesimals and determine the form of (a maximum of two) 
indeterminate functions possibly contained in e1 and/or e2, such as to 
solve the problem.

\item \verb-HINT=[[e1,e2], [e3,e4],...]-, where \verb-[e1,e2]- is any of 
the above.

\item \verb-HINT=parametric-, indicates to the solver that it should only 
look for a parametric solution for the given (1\st order) ODE.

\end{itemize}

Finally, there are three global variables managing the solving process, 
which are automatically set by internal routines but can also be assigned 
by the user, as desired. They are {\it dgun}, {\it ngun}, {\it sgun}, for 
setting, respectively, the maximum {\it d}egree of polynomials entering 
some of the ans\"atze for the infinitesimals, the maximum {\it n}umber of 
subproblems into which the original ODE should be mapped, and the maximum 
{\it s}ize permitted for such subproblems. The dgun variable is 
automatically set each time symgen is called, according to the given ODE, 
whereas, by default, ngun and sgun have their values assigned to 1. 
Increasing the value of {\it d}gun usually helps, especially in the case 
of polynomial ODEs; but, although the user-assigning of the {\it n}gun or 
{\it s}gun variables might increase the efficacy of the algorithms, each 
increase of one unit can slow down the solving process geometrically.

\section{Finding the {\it infinitesimals}}
\label{algorithms}

In the context of SM, the infinitesimals we are looking for are solutions 
of \eq{xi_eta}, which is linear in the functions $(\xi, \eta)$ and their 
derivatives. The command which looks for these pairs of infinitesimals for 
a given ODE is {\bf symgen}, and the version here presented is an 
extension to 2\nd order ODEs of the schemes presented in \cite{odetools1}.

\subsection{The {\bf symgen} subroutines}

The key idea underlying this extension is that to look for {\it dynamical} 
symmetries for 2\nd order ODEs is mainly equivalent to looking for {\it 
point} symmetries for 1\st order ODEs; except that the routines are now 
going to look for solutions involving $(x,y,\y1)$ instead of only $(x,y)$. 
More specifically, we adapted the previous {\bf symgen/...} subroutines, 
whose main purpose is to look for particular solutions to linear problems, 
in order to look for such particular solutions considering $\y1$ as a new 
variable in equal footing as $x$ and $y$. An explanation with details 
about how these {\bf symgen/...} routines work can be found in 
\cite{odetools1}; so, we here restricted the discussion to some examples 
illustrating the type of results which can now be obtained.

The $1^{st}$ algorithm, \abaco{1}, typically looks for infinitesimals of
the form $[\xi=0,\eta=F(q)]$ and $[\xi=F(q),\eta=0]$, where q is one of
$(x,y,\y1)$.



\ni {\it Example:} Kamke's 2\nd order non-linear ODE 206


\begin{equation}
\ode
\l({a}^{2}-{x}^{2}\r)\l({a}^{2}-y^{2}\r)
{\frac {d^{2}y}{d{x}^{2}}}
+\l({a}^{2}-{x}^{2}\r)y\l({\frac {dy}{dx}}\r)^{2}
-x\l({a}^{2}-y^{2}\r){\frac {dy}{dx}}=0
\label{abaco1_ODE}
\end{equation}
\begin{verbatim}
> symgen(ode,way=abaco1);
\end{verbatim}
\begin{eqnarray}
\lefteqn{[\xi=0,\ 
\eta=\sqrt {-{a}^{2}+{y}^{2}}\left (1+\ln (y+\sqrt {-{a}^{2}+{
y}^{2}})\right )],}
& & 
\nonumber \\
& &
[\xi=\sqrt {{x}^{2}-{a}^{2}}\left (1+\ln (x+\sqrt {{
x}^{2}-{a}^{2}})\right ),\ \eta=0]
\end{eqnarray}
It is also not difficult to find the patterns of 2\nd order ODEs having
the type of symmetries which \abaco{1} is prepared in principle to look
for. These ODE-patterns can be obtained using the {\bf equinv} routine,
programmed to solve the inverse problem ({\it input}=symmetries,
{\it output}= invariant ODE; see \se{inverse}). For example, in the first
of the six cases mentioned above,

\begin{verbatim}
> equinv([F(x),0],y(x),2);   
\end{verbatim}
\begin{equation}
{\frac {d^{2}y}{d{x}^{2}}}=
\frac{1}{\left (F(x)\right )^{2}}
\l(
{{{\it \_F1}\l(y,F(x){
\frac {dy}{dx}}\r)}}
-{{{\frac {dF(x)}{dx}}}{F(x)}{\frac {dy}{dx}}}
\r)
\end{equation} 
is the most general 2\nd order ODE invariant under $[F(x),0]$, $F$ and
{\it \_F1} being arbitrary functions of its arguments, and \eq{abaco1_ODE}
is a particular case of this ODE family.

The $2^{nd}$ and $3^{rd}$ algorithms look for polynomial in
$(x,y,\y1)$ solutions to \eq{xi_eta}.

\ni {\it Example:} Kamke's 2\nd order non-linear ODE 181


\begin{equation}
\ode
{x}^{2}\l(x+y\r)\ {\frac {d^{2}y}{d{x}^{2}}}
-\l(x{\frac {dy}{dx}}-y\r)^{2}=0
\end{equation}

\begin{verbatim}
> symgen(ode,way=3);
\end{verbatim}
\begin{equation}
[\xi=x,\eta=y],\ \ \ 
[\xi=-x,\eta=x],\ \ \ 
[\xi={x}^{2},\eta=xy]
\end{equation}

\begin{verbatim}
> odsolve(ode);
\end{verbatim}
\begin{equation}
y=x{e^{\l(^{ -{\frac {C_1}{x}}+C_2}\r)}}-x
\end{equation}

The $4^{th}$ algorithm looks for rational in $(x,y,\y1)$ solutions to
\eq{xi_eta}.

\ni {\it Example}

%
%
%


\begin{equation}
\ode
{\frac {d^{2}y}{d{x}^{2}}}=
\frac{1}{x+y}
{\left (2\,{\frac {dy}{dx}}+1\right )
{\frac {dy}{dx}}}
\end{equation}

\begin{verbatim}
> symgen(ode,way=4);
\end{verbatim}
\begin{eqnarray}
\lefteqn{[[\xi=-\frac{1}{x+y},\eta=0],\ \
[\xi=-{\frac{y}{x+y}},\eta=0],\ \ 
[\xi={\frac {x\l(x+2\,y\r)}{x+y}},\eta=0],}
& &
\nonumber
\\*[.25 in]
&
[\xi=-1,\eta=1],\ \
[\xi=x,\eta=y]]
&
\end{eqnarray}

\begin{verbatim}
> odsolve(ode,way=4,can2);
\end{verbatim}
\begin{equation}
y=-{\frac {{x}^{2}+2\,C_{1}}{2\,x+2\,C_{2}}}
\end{equation}

The $5^{th}$ algorithm uses a polynomial of degree two constructed from a
basis of functions and algebraic objects, together with their derivatives,
taken from the given ODE (see \cite{odetools1}).

\ni {\it Example:} Kamke's 2\nd order non-linear ODE 238


\begin{equation}
\ode
2\,\l(1+{x}^{2}\r)\l({\frac {d^{2}y}{d{x}^{2}}}\r)^{2}
- x\l(x+4\,{\frac{dy}{dx}}\r)
{\frac {d^{2}y}{d{x}^{2}}}
+ 2\,\l(x+{\frac {dy}{dx}}\r){\frac{dy}{dx}}-2\,y=0
\end{equation}
\begin{verbatim}
> symgen(ode,way=5);
\end{verbatim}
\begin{equation}
[\xi=0,\eta={\frac {4\,\y1+2\,x+{x}^{3}-x\sqrt
{{x}^{4}-8\,{x}^{3}\y1-16\,x\,\y1-
16\,{\y1}^{2}+16\,y+16\,{x}^{2}y}}{4+4\,{x}^{2}}}]
\end{equation}

\ni In the example above, the square root, which appeared after isolating
$ \displaystyle {\frac {d^{2}y}{d{x}^{2}}}$, was viewed as an {\it
algebraic object}, and the symmetry found is of dynamical type.

The next algorithm, \abaco{2}, looks for infinitesimals of the form
$[\xi=F(q_1),\eta=G(q_2)]$, where $q_1$ and $q_2$ are members of
$\{x,y,\y1\}$.

\ni {\it Example}


\begin{equation}
\ode
{\frac {d^{2}y}{d{x}^{2}}}=\left ({\frac {dy}{dx}}\right )^{2}
-
\frac{e^{y}}{x^{2}}\
F\l(\frac{1}{e^{y}}+\ln (x),
{\frac {x}{e^{y}}\l({\frac {dy}{dx}}\r)}\r)
\end{equation}
\begin{verbatim}
> symgen(ode,way=abaco2);
\end{verbatim}
\begin{displaymath}
[\xi=x,\ \eta=e^{y}]
\end{displaymath}


Apart from setting up the {\bf symgen/...} routines to work with 2\nd
order ODEs, two new algorithms for 1\st order ODEs were implemented. The
first one is specific for Riccati-type ODEs and the second one is based on
a direct attempt to solve the problem using the {\bf pdsolve} routine from
the {\it PDEtools} package.
%
%
%
%
The routine for Riccati ODEs makes three sequential attempts to solve
\eq{xi_eta}, the key idea being to map the problem of finding the
infinitesimals for a given Riccati ODE into that of finding particular
solutions to a related linear 2\nd order PDE for $\eta(x,y)$. These three
attempts consist of tackling this PDE by: using the $5^{th}$ algorithm
explained above, trying separation of variables, and using the \abaco{1}
scheme.

\ni {\it Example:} Kamke's 1\st order ODE 16


\begin{equation}
\ode
{\frac {dy}{dx}}+y^{2}+\l(x\,y-1\r)f(x)=0
\end{equation}
\begin{verbatim}
> symgen(ode,way=6);
\end{verbatim}
\begin{equation}
[\xi=0,\eta= {e^{^{\l(
\displaystyle
\int
{\frac {f(x)x^2+2}{x}} {dx}\r)}}}
{\frac {\l(y\,x-1\r)^{2}}{x^2}}]
\end{equation}
\begin{verbatim}
> odsolve(ode,way=6);
\end{verbatim}
\begin{equation}
y=
\frac{1}{x}
+
\frac
{{e^{^{\l(-
\displaystyle
\int \!
f(x)x{dx}-2\,\ln (x)\r)}}}}
{
\displaystyle
\int \!{e^{^{\l(-
\int \!{
\frac {f(x){x}^{2}+2}{x}}{dx}\r)}}}{dx}-C_{1}}
\end{equation}

\subsection{The inverse problems}
\label{inverse}
For many reasons, it appeared interesting to develop routines to solve the
inverse problems too; that is: find the most general 1\st/2\nd order ODE
simultaneously invariant under a given set of symmetries, or given the
answer of an {\it unknown} ODE, find its symmetries (and then the
corresponding invariant ODE). Typical applications for these routines are:
we look for a description of a problem for which we only know the symmetries
or some particular solutions; or we look for solving methods for ODEs having
a given pattern, for which we know the solution in some cases, and we want
to find the solution for the general case.

The {\it ODEtools} routines related to solving the inverse problems are
{\bf equinv} and {\bf buildsym}. The former is prepared to work, in
principle, with an ``arbitrary number" of symmetries. As a first example,
consider the determination of the most general 2\nd order ODEs
simultaneously invariant under

\begin{equation}
\X
[[\xi=1,\,\eta=1],[\xi=x,\,\eta=y],[\xi=x^2,\,\eta=y^2]]
\end{equation}
\begin{verbatim}
> equinv(X, y(x), 2);
\end{verbatim}

\begin{equation}
{\frac {d^{2}y}{d{x}^{2}}}=\l(2\,\sqrt {{\frac {dy}{dx}}}
+{\frac{2}{\sqrt {{\frac {dy}{dx}}}}}
+C_{1}\r)\l({\frac {dy}{dx}}\r)^{3/2}
\frac {1}{\l(y-x\r)}
\end{equation}

As an example involving dynamical symmetries, consider the determination of
the 2\nd order ODE simultaneously invariant under

\begin{equation}
\X
[[\xi=\y1,\,\eta=0],\ 
[\xi=0,\,\eta=\frac{1}{\y1}]];
\end{equation}

\begin{verbatim}
> equinv(X, y(x), 2);
\end{verbatim}
\begin{equation}
{\frac {d^{2}y}{d{x}^{2}}}=
-{
\frac {
\displaystyle
\l({\frac {dy}{dx}}\r)^{2}}
{
\displaystyle
-2\,x\,{\frac {dy}{dx}}
+2\,y-{\it \_F1}\l({\frac {dy}{dx}}\r)\l({\frac {dy}{dx}}\r)^{2}}}
\end{equation}

where ${\it\_F1}$ is an arbitrary function of its argument. Proceeding as in
the examples above, it is possible to determine the general ODE-patterns
which can be solved by the \abaco{1} scheme and some of those which can be
solved by \abaco{2}, as well as to build matching-pattern routines to match
families of ODEs for which we know the form of the symmetries {\it a
priori}. Of course, since the number of possible ODE families is infinite,
it is not realistic to think of producing matching pattern routines for all
the cases. On the other hand, for a variety of concrete problems which arise
in different applied areas, it may be convenient to build these routines
and have the related solving schemes at hand. Remarkably, most of the
standard methods actually work in this manner (i.e., by matching a pattern
and then applying the corresponding solving scheme), and these patterns
actually arose from applied problems. As a benchmark for {\bf equinv}, the
routine succeeded in finding the most general invariant ODEs for all the
twenty examples found in sec. 8.3 or Ibragimov's book \cite{ibragimov} on
Lie methods.

Concerning the computational scheme used in {\bf equinv}, the idea is,
roughly speaking, to introduce the given symmetries into \eq{xi_eta}, and
solve it for the right-hand-side of the ODE, which is now viewed as ``the
unknown of the problem". One of the interesting advantages of this simple
approach is that we can consider the invariance of an ODE under point and
dynamical symmetries in equal footing, and forget about the technicalities
involved in defining differential invariants when derivatives are present in
the transformation equations\footnote{This scheme works just as well to
determine high order ODEs (differential order $ > 2$) invariant under
dynamical symmetries.}.

When many symmetries are given at once, from the computational point of
view, the problem is a bit more complicate. A brief summary of the
scheme implemented is as follows:

\begin{enumerate}

\item introduce the first pair of infinitesimals into \eq{xi_eta} setting up
a PDE for the right-hand-side (RHS) - here denominated $\Phi(x,y,\y1)$ - of
the most general explicit invariant second order ODE;\label{set_PDE1}

\item solve this PDE for $\Phi$; the solution will involve an arbitrary
function of two differential invariants, say
$F(\psi_1(x,y,\y1),\psi_2(x,y,\y1))$;

\item update the symbolic value of $\Phi$ introducing the result of the
previous step (i.e., introducing $F$). This updated $\Phi(x,y,\y1)$ is an
explicit expression of $\{x,\ y,\ \y1\}$ hereafter called $\phi(x,y,\y1)$.
When only one symmetry is given, $\phi(x,y,\y1)$ is already the RHS of the
invariant ODE we are looking for;\label{set_phi}

\item take \eq{xi_eta} again and set up a new PDE using the second given
pair of infinitesimals, and {\it substitute} $\Phi$ found in PDE by its
updated value $\phi$;\label{recurse}

\item we now need to solve this new PDE for the arbitrary function $F$
introduced in step (\ref{set_phi}), but this is not directly possible
since the unknown $F$ depends on {\it algebraic} expressions
- $\psi_1(x,y,\y1)$ and $\psi_2(x,y,\y1)$; so introduce these differential
invariants as new variables to obtain a PDE for $F(\psi_1,\psi_2)$ where
$(\psi_1,\psi_2)$ are now {\it symbol} variables;\label{set_PDE2}

\item solve the PDE of the previous step for $F$ and use the inverse
transformation to restore the original variables $\{x,\ y,\ \y1\}$ in the
answer;

\item update the symbolic value of $\Phi$ with respect to the result of
the previous step. The resulting ODE $y^{''} = \phi(x,y,\y1)$ is now
simultaneously invariant under the first two pairs of infinitesimals;

\item if more pairs of infinitesimals are given, repeat this process from
step (\ref{recurse});

\end{enumerate}

To solve all the intermediate PDEs we use the {\bf pdsolve} command of the
PDEtools package \cite{pdetools}, with which ODEtools is completely
integrated\footnote{In turn, the solving of intermediate ODEs built by {\bf
pdsolve} to solve PDEs is carried up by ODEtools commands. PDEtools and
ODEtools are both compiled in a single binary maple library.}. What requires
a bit more of care is the determination and handling of all the changes of
variables involved at each step. 

Concerning the {\bf buildsym} command, the idea is to build the
symmetries of an {\it unknown} ODE departing from its solution. When the
command succeeds it is then possible to use the {\bf equinv} command to
build the most general ODE having the symmetries of the given ODE
solution. As a simple example, consider again \eq{linear_ODE}. Passing the
answer of this ODE (\eq{ans_linear_ODE}) to {\bf buildsym} it returns two
symmetries:

\begin{verbatim}
> X := buildsym(ans, y(x));
\end{verbatim}
\begin{equation}
{\rm X := }[\xi=0,\ \eta=x],\ \ [\xi=0,\ \eta=x\int \!{
\frac {{e^{\int \!{\it \_F1}(x){dx}}}}{{x}^{2}}}{dx}]
\end{equation}

\ni In this case, since the ODE is linear and the symmetries are of the
form $[0,F(x)]$, the returned symmetries are actually two particular
solutions to \eq{ans_linear_ODE}. A less simpler example is given by:

\begin{equation}
{\frac {d^{2}y}{d{x}^{2}}}=
{\frac {{x}^{2}}{\sqrt {y}
}}
+{\frac {dy}{dx}}
-\frac{1}{2\,y}\,{{\left ({\frac {dy}{dx}}
\right )^{2}}}
\label{nonlinear_ODE}
\end{equation}

\ni An implicit solution to this ODE is given by:

\begin{equation}
\frac{4}{3}\,y^{3/2}+\frac{2}{3}\,{x}^{3}+2\,{x}^{2}+4\,x-2\,C_1\,{e
^{x}}-C_2=0
\end{equation}

\ni This ODE solution leads to the following pair of underlying symmetries
\begin{verbatim}
> X := buildsym(ans, y(x));
\end{verbatim}
\begin{equation}
{\rm X := }
[\xi=0,\ \eta={\frac {1}{\sqrt {y}}}],\ \ 
[\xi=0,\ \eta={\frac {{e^{x}}}{\sqrt {y}}}]
\end{equation}

\ni In turn this information leads to the general ODE family having
a solution with the same two symmetries of \eq{nonlinear_ODE}:
\begin{verbatim}
> equinv(X, y(x));
\end{verbatim}
\begin{equation}
{\frac {d^{2}y}{d{x}^{2}}}=
{\frac {{\it \_F1}(x)}{\sqrt {y}
}}
+{\frac {dy}{dx}}
-\frac{1}{2\,y}\,{{\left ({\frac {dy}{dx}}
\right )^{2}}}
\end{equation}

Concerning the computational scheme implemented in {\bf buildsym}, the
idea is, roughly speaking, to use the ODE-solution to build two {\it first
integrals}, from which two pairs of infinitesimals can be obtained. To
build these pairs of infinitesimals, we note that since the first
integrals, say $\varphi$ and $\psi$, are solutions of the linear operator

\begin{equation}
{\rm A := } f \rightarrow \frac{\partial f}{\partial x}
+ \y1 \frac{\partial f }{\partial y}
+ \Phi(x,y,\y1)
\frac{\partial f}{\partial \y1}
\label{A}
\end{equation}

\ni and from the definition of a symmetry $X$ such that it satisfies
$[X,A] = \lambda A$ (see \cite{stephani}) it is always possible to
formulate the problem as: to determine two infinitesimal generators $X_1$
and $X_2$ of the form $[0,\ \eta]$ satisfying

\begin{eqnarray*}
X_{1}{\varphi_{1}}=\eta_{1} \frac{d\varphi}{dy}
+\eta_{1}^{(1)}\frac{d\varphi}{d\y1}=0,\:\:\:\:\:\:
X_{1}{\varphi_{2}}=\eta_{1}\frac{d\psi}{dy}
+\eta_{1}^{(1)}\frac{d\psi}{d\y1}=1,\\*[.25in]
X_{2}{\varphi_{1}}=\eta_{2}\frac{d\varphi}{dy}
+\eta_{2}^{(1)}\frac{d\varphi}{d\y1}=1,\:\:\:\:\:\:
X_{2}{\varphi_{2}}=\eta_{2}\frac{d\psi}{dy}
+\eta_{2}^{(1)}\frac{d\psi}{d\y1}=0. 
\end{eqnarray*}

This system of equations can be solved for $\{\eta_{1}\
\eta_{2}\}$, arriving at

\begin{eqnarray}
\eta_{1} = {\frac
{\left|\begin{array}{cc}
0 & \frac{d\varphi}{d\y1} \\
1 & \frac{d\psi}{d\y1} \\
\end{array}\right|}
{\left|\begin{array}{cc}
\frac{d\varphi}{dy} & \frac{d\varphi}{d\y1} \\
\frac{d\psi}{dy} & \frac{d\psi}{d\y1} \\
\end{array}\right|}}
\:\:\:\:
\eta_{2} = {\frac
{\left|\begin{array}{cc}
1 & \frac{d\varphi}{d\y1} \\
0 & \frac{d\psi}{d\y1} \\
\end{array}\right|}
{\left|\begin{array}{cc}
\frac{d\varphi}{dy} & \frac{d\varphi}{d\y1} \\
\frac{d\psi}{dy} & \frac{d\psi}{d\y1} \\
\end{array}\right|}}
\end{eqnarray}

The determination of $X_1$ and $X_2$ then amounts to the determination of
$\varphi$ and $\phi$. These first integrals can be determined from the
given solution as follows:

\begin{enumerate}

\item solve the given ODE solution with respect to one of the
integration constants, say $C_2$, obtaining $C_2=F(x,y(x),C_1)$;

\item differentiate the result of the previous item w.r.t $x$ and solve it
w.r.t $C_1$ obtaining the first integral (a solution of \eq{A})
$C_1=\varphi(x,y,\y1)$;

\item repeat the two previous items interchanging $C_1 \leftrightarrow
C_2$ to obtain the second first integral $C_2=\psi(x,y,\y1)$;

\end{enumerate}

\section{Integrating the ODE after finding the infinitesimals}
\label{integrating}

The methods implemented for integrating the ODE once the infinitesimals
were found (methods (a), (b) and (c) mentioned in \se{liemethod}) are
extensions to those found in \cite{stephani,bluman,olver}. The extension
consists of making it possible to use them to integrate the given ODE even
when the symmetries found are of dynamical type. Since their use in the
framework of point symmetries is rather standard, we restricted the
discussion here to the extension of these schemes to work with dynamical
symmetries.

Concerning ``why" so many integration strategies were implemented, we
would like to recall that the knowledge of 2 symmetries of a 2\nd order
ODE does not guarantee a successful integration of it since the different
methods available do not apply in {\it all} the cases. Also, two methods
which may be ``equivalent" from the mathematical point of view, may not be
equivalent for the computer algebra system, since the intermediate tasks
involved in each integration process are different. We then preferred to
implement all these methods, provide a {\it default} for their use, and
make them available to the user by giving extra arguments to {\bf odsolve}
(see \se{description}).

\subsection{Using differential invariants}
\label{dif}

As is well known, once a pair of infinitesimals for a given 2\nd order ODE
is found, it is possible to integrate the ODE if one succeeds in using
these infinitesimals to find the differential invariants of order 0 and 1
and in solving an intermediate 1\st order ODE (see for instance section
3.3.2 of \cite{bluman}). However, if the symmetry is of dynamical type,
the invariant of order 0 will be a function of \y1, the invariant of order
1 will be a function of \yy2, and so on. Then, the whole scheme breaks down
since, to reduce the order, the scheme requires that the differential
invariants of order {\it n} do not depend on derivatives of order higher
than {\it n}.

The alternative we implemented is based on the observation that, if we
transform the symmetry to the evolutionary form\footnote{Since
$[\xi=1,\eta=\y1]$ is always a (trivial) symmetry, a general symmetry can
always be ``gauged", that is, rewritten in many different forms. The
evolutionary form is given by $[\xi=0,\eta=\Phi(x,y,\y1)]$.} \cite{olver},
then $x$ will always be an invariant (of order 0), so it is possible to work
as follows:

\begin{enumerate}

\item  put the symmetry in the evolutionary form;

\item use $u=x$ as the invariant of order 0 and find the invariant of order
1, $v=v(x,y,\y1)$;

\item  construct the invariant of order 2: $(\frac{dv}{du})$;

\item change the variables and write the ODE in terms of $u$, $v$ and
$\frac{dv}{du}$ obtaining the expected reduction of order;

\item solve this 1\st order ODE, and reintroduce the original variables
$(x,y,\y1)$, obtaining another 1\st order ODE, which can be integrated
straightforwardly since it has the same symmetry of the original 2\nd order
ODE.

\end{enumerate}  
A tricky point in this last step is that, when the symmetry is of
dynamical type (depends on \y1), to obtain the symmetry which holds for
the 1\st order ODE one should replace all occurrences of \y1 in the
infinitesimals of the original ODE by the right-hand-side of this 1\st
order ODE.

\subsection{Using first integrals}
\label{gon}

Another way of integrating a 2\nd order ODE is to use 2 pairs of
infinitesimals to build 2 first integrals, from which one can expect to
obtain a closed form solution (see section 7.3 of \cite{stephani}). One of
the advantages of this method is that one always work with the {\it
original variables} avoiding potential problems related to the
introduction of changes of variables (e.g., the system fails in finding
the inverse transformation etc.). Taking into account the way we use
``dynamical symmetries" to solve the 1\st order ODE obtained in the
reduction process (see \se{dif}), the extension of this scheme to work
with such symmetries was straightforward.

Now, it is also possible to tackle a 2\nd order ODE from the knowledge of
only 1 pair of infinitesimals (either of point or dynamical type), by
looking for first integrals, exploiting similar ideas as those of the
previous subsection; that is:

\begin{enumerate}
\item[(i)] Put the known symmetry in the evolutionary form.
\label{item_1};
\end{enumerate}

One can then prove that there exists a first integral
$\varphi(x,y,\y1)$ which satisfies:
\begin{eqnarray}
A(\varphi) & = & 0
\nonumber
\\
X(\varphi) & = & 0
\label{A_X_0}
\end{eqnarray}
where $A$ is given by \eq{A}, and $X$ is the infinitesimal symmetry generator
\begin{displaymath}
X := f \rightarrow \xi(x,y,\y1)\,\frac{\partial f}{\partial x}
+ \eta(x,y,\y1)\, \frac{\partial f}{\partial y}
+ \eta_1(x,y,\y1)\,\frac{\partial}{\partial \y1}
\end{displaymath}
\begin{enumerate}
\item[(ii)] Solve the characteristic strip of $X(\varphi)=0$ to obtain the 
invariants of order 0 and 1: $c_0(x,y,\y1)$ and $c_1(x,y,\y1)$.
\end{enumerate}
Since the symmetry is in evolutionary form, it always happens that $c_0=x$,
and the characteristic strip is reduced to a single 1\st order ODE. Now,
observing that the solution to $X(\varphi)=0$ can always be written as
$\varphi(c_0,c_1)$,

\begin{enumerate}
\item[(iii)] change the variables in $A(\varphi)=0$, from $(x,y,\y1)$ to
$(c_0,y,c_1)$.
\end{enumerate} 
One can then prove that $A(\varphi)=0$ becomes
$$\l(\frac{\partial}{\partial c_0}
+ A(c_1)\frac{\partial}{\partial c_1}\r)\varphi(c_0,c_1)=0$$

that is, a PDE in only two variables; solve it, obtaining $\varphi(c_0,c_1)$
now simultaneously satisfying equations (\ref{A_X_0}). Finally,

\begin{enumerate} \item[(iv)] reintroduce the original variables to obtain
$\varphi(x,y,\y1)$; that is, a 1\st order ODE which can be integrated
straightforwardly since it already satisfies $X(\varphi)=0$. \end{enumerate}

\subsection{Using canonical coordinates}
\label{can}

It is also possible to use a pair of infinitesimals to constructively
reduce the order of a 2\nd order ODE by introducing the canonical
variables of the related invariance group. Also, if one knows two or more
symmetries at once, one can use this information to integrate the ODE in
almost only one step (see sec. 7.4 of \cite{stephani}). These two methods
were implemented too, although their extension to work with dynamical
symmetries doesn't appear to us as advantageous if compared to the
extensions outlined in the previous subsections.

\subsection{Examples}

We present here some examples to illustrate how these integration schemes
work and the different types of answers they may return. As the first
example, consider Kamke's non-linear 2\nd order ODE 99:

\begin{equation}
\ode
{x}^{4}{\frac {d^{2}y}{d{x}^{2}}}+\l(x{\frac {dy}{dx}}-y\r)^{3}=0
\end{equation}
This ODE has simple polynomial point symmetries:

\begin{verbatim}
> symgen(ode);
\end{verbatim}
\begin{equation}
[\xi=0,\eta=x],\ \ \ [\xi=x,\eta=y]
\end{equation}
Its integration is rather fast using the method explained in \se{gon}:

\begin{verbatim}
> odsolve(ode,gon);
\end{verbatim}
\begin{equation}
y=\l(-\arctan\l({\frac {1}{\sqrt {-1+{x}^{2}C_{1}}}}\r)+
C_{2}\r)x
\label{esta}
\end{equation}
Remarkably, if in this case we use the integration scheme which uses the 2
symmetries to build 2 first integrals, the answer will not be so compact
due to the manipulations required to transform the two first integrals
$\varphi_1(x,y,\y1),\ \varphi_2(x,y,\y1)$ into the solution
$\varphi_3(x,y)$. The opposite happens with Kamke's ODE 227:

\begin{equation}
\ode
\l(x{\frac {dy}{dx}}-y\r){\frac {d^{2}y}{d{x}^{2}}}
+4\,\l({\frac {dy}{dx}}\r)^{2}=0
\end{equation}
for which {\bf symgen} finds two point symmetries, $[\xi=x,\eta=0]$ and
$[\xi=0,\eta=y]$, and the more compact answer is returned, using special
functions, by this scheme which builds two first integrals:

\begin{verbatim}
> odsolve(ode,gon2);
\end{verbatim}
\begin{equation}
y=\frac{1}{2}\,\l({\rm LambertW}\l(-\frac{x}{e^{^{\l(2+C_{2}\r)}}}\r)+2\r)
{e^{-\l({\rm LambertW}\l(-\frac{x}{e^{^{\l(2+C_{2}\r)}}}\r)+2\r)}}
{e^{C_{1}}}
\end{equation}

As the next example, consider Kamke's ODE 122:

\begin{equation}
\ode
\l({\frac {d^{2}y}{d{x}^{2}}}\r)y
-\l({\frac {dy}{dx}}\r)^{2}-f(x)\,y \l({\frac {dy}{dx}}\r)-g(x)\,y^{2}=0
\end{equation}
This example involves two arbitrary functions of $x$ and the package
defaults only succeed in finding 1 symmetry:

\begin{verbatim}
> symgen(ode);
\end{verbatim}
\begin{equation}
[\xi=0,\eta=y]
\end{equation}
In this situation only the methods prepared to work with just 1 symmetry
will work; in this example, all of them lead to the same result:
\begin{verbatim}
> odsolve(ode,gon);
\end{verbatim}
\begin{equation}
y={e^{^{
\displaystyle
\l(\int \!
{e^{^{\int \!f(x){dx}}}}\int \!g(x){e^{^{\l(-\int \!f(x){
dx}\r)}}}{dx}+{e^{^{\int \!f(x){dx}}}}C_{1}{dx}+C_{2}\r)}}}
\end{equation}

The following example, Kamke's ODE number 20, is rather critical in the
sense that all the integration schemes fail in solving the problem, except
the least elaborated one (constructive reduction of order introducing
canonical coordinates, using only 1 symmetry at a time). Due to the presence
of an arbitrary function of $x$ and $y(x)$, the answer appears in implicit
form:

\begin{equation}
\ode
{\frac {d^{2}y}{d{x}^{2}}}=
{x}^{(-\frac{3}{2})}\,F\l({\frac {y}{\sqrt {x}}}\r)
\end{equation}

\begin{verbatim}
> odsolve(ode,can);
\end{verbatim}
\begin{equation}
\frac{1}{2}\,\ln(x)
-\int^{
\displaystyle
\frac{y}{\sqrt {x}}}
{\frac {1}{\sqrt {{{\it \_b}}^{2}
+ 8\,
\displaystyle
\int \!F(\_b)\,{d\_b}+ C_{1}}}
}\,d\_b
-C_{2}=0
\end{equation}

As the last example, consider the case in which only 1 symmetry has been
found, and the integration schemes only succeed in reducing the order but
not in solving the resulting 1\st order ODE; e.g., Kamke's ODE number 225:

\begin{equation}
\ode
F(y){\frac {d^{2}y}{d{x}^{2}}}-\mbox {D}(F)(y)\l({\frac {dy}{dx}}\r)^{2}
-\l(F(y)\r)^{2}\Phi\l(x,{\frac {1}{F(y)}}\l({\frac {dy}{dx}}\r)\r)=0
\end{equation}
A symmetry for this ODE is found by the fifth algorithm,
\begin{verbatim}
> symgen(ode,way=5);
\end{verbatim}
\begin{equation}
[0,F(y)]
\end{equation}
In these cases, an answer with the structure of the solution is returned:

\begin{verbatim}
> odsolve(ode,way=5);
\end{verbatim}
\begin{eqnarray}
y & = & {\rm RootOf}\l(C_1+\int \!{\it \_s1}({\it \_r}){d{
\it \_r}}-\int ^{{\it \_Z}}\! \frac{1}{F({\it \_a})}{d{
\it \_a}}\r)
\nonumber
\\*[.2 in]
 & & {\rm \&where}
\l[\left \{{\frac {d}{d{\it \_r}}}{\it \_s1}({\it \_r})=\Phi(
  {\it \_r},{\it \_s1}({\it \_r}))\right \},
\left \{{\it \_r}=x,\ {\it 
\_s1}({\it \_r})={\frac {1}{{F(y)}}{\frac {dy}{dx}}}\right \}\r]
\end{eqnarray}

In the structure above, we see the answer for $y$ up to a change of
variables. More specifically, this change of variables was not performed
because the system doesn't know how to solve the displayed intermediate
1\st order ODEs. Although in this particular case it is impossible to go
ahead -the unsolved 1\st order ODE is the most general we can imagine - in
many cases the user may be able to solve the reduced first order ODE by
other means. Once ${\it \_s1}({\it \_r})$ is obtained, a change of
variables using the displayed transformation, will lead to the desired
answer for $y$.

\section{Tests and performance}
\label{tests}

We tested the set of routines here presented, mainly to confirm the
correctness of the returned results. In addition, we ran a {``\it
performance test"} of {\bf symgen} determining the infinitesimals for
Kamke's set of 246 non-linear 2\nd order examples. Also, although the
primary goal of {\bf odsolve} is to complement the standard Maple {\bf
dsolve} in solving {\it non-classifiable} ODEs, we ran a {``\it comparison
of performances"} in solving these Kamke examples too. The idea underlying
both performance tests was to see the ability of the {\bf symgen} command in
determining infinitesimals for a well known set of equations, and to compare
the possible efficacy of a {\it SM-based} solver with that of a {\it
classification-based} solver such as {\bf dsolve}. 

\subsection{Test of {\bf symgen}}

As explained in sec.\ref{liemethod}, the SM approach for solving ODEs
involves two main steps: the determination of pairs of infinitesimals and
their subsequent use in the integration process. The first {\it
performance} test, thus, concerned the explicit determination, by the {\bf
symgen} command, of infinitesimals for the above mentioned 246 non-linear
examples of Kamke's book. To perform these tests we prepared 3 files
containing the input of all these ODE examples. The table below displays
the total number of successes, the average time spent with each
solved/fail case, and the number of successes of {\it each} of {\bf
symgen}'s six algorithms when the whole set of ODEs was tackled using only
one of them\footnote{The {\it infinitesimals} found by {\bf symgen} as
well as the answers found by {\bf odsolve} for Kamke's non-linear 2\nd
order ODEs are available at: http://dft.if.uerj.br/odetools.html}. 

{\begin{center} {\footnotesize
\begin{tabular}
{|c|c|c|c|c|c|c|c|c|c|c|c|}
\hline
 & & \multicolumn{2}{|c|}{Symmetries} & \multicolumn{2}{|c|}{Average time} &
\multicolumn{6}{|c|}{infinitesimals determined by} \\
\cline{3-12}
File & ODEs & only 1  & many & {\it ok} & {\it fail}& \abaco{1} & 2 & 3 & 4 
& 5 & \abaco{2} \\
\hline 
1 &  100 & 49 & 24 &  7 sec. &  22 sec. & 54 & 62 & 72 & 65 & 55 & 71 \\
\hline 
2 &  100 & 26 & 53 &  6 sec. &  22 sec. & 67 & 53 & 74 & 68 & 56 & 70 \\
\hline 
3 &  46 & 17 & 22 &  26 sec. &  54 sec. & 29 & 20 & 30 & 27 & 23 & 31 \\
\hline 
\hline 
Total: & 246 & 92 & 99 & $ \approx 10$ sec. & $\approx 29$ sec.
                                      & 150 & 138 & 176 & 160  & 134 & 171\\
\hline 
\multicolumn{12}{c}
{Table 1. \it Kamke's non-linear 2\ind order ODEs: Infinitesimals determined 
by {\bf symgen}.}\\
\end{tabular} }
\end{center}}
We here distinguished the cases in which only one symmetry has been found,
since in those cases the integration procedure has greater chances of only
obtaining a reduction of order. The Kamke ODE numbers related to the
table above are:
{\begin{center} {\footnotesize
\begin{tabular}{|p{1.2in}|p{4.8in}|}
\hline
{\bf symgen} found  &  Kamke's book ODE-number\\
\hline 
only 1 symmetry for 92 ODEs &
11, 12, 15, 17, 20, 21, 22, 23, 24, 25, 26, 28, 31, 32, 40, 45, 46, 47, 48,
49, 54, 58, 64, 65, 66, 68, 69, 70, 72, 73, 74 , 75, 76, 77, 79, 80, 82, 83,
86, 87, 89, 90, 91, 92, 94, 96, 97, 98, 100, 102, 103, 105, 106, 118, 119,
120, 121, 122, 129, 132 , 152, 153, 156, 160, 165, 170, 172, 174, 177, 183,
186, 187, 196, 197, 200, 201, 202, 203, 204, 206, 208, 213, 219, 223, 224,
225, 226, 230, 231, 238, 241, 242
   \\
\hline
many symmetries for 99 ODEs &
1, 2, 4, 7, 10, 14, 30, 42, 43, 50, 56, 57, 60, 61, 62, 63, 67, 71, 78, 81,
84, 88, 93, 99, 104, 107, 108, 109, 110, 111, 113, 117, 124, 125, 126, 127,
128, 130, 133, 134, 135, 136, 137, 138, 140, 141, 143, 146, 150, 151, 154,
155, 157, 158 , 159, 162, 163, 164, 166, 168, 169, 173, 175, 176, 178, 179,
180, 181, 182, 184, 185, 188, 189, 190, 191, 192, 193, 205, 209, 210, 214,
218, 220, 221, 222, 227, 228, 229, 232, 233, 234, 236, 237, 239, 240, 243,
244, 245, 246
   \\
\hline
\multicolumn{2}{c}{Table 2. \it
Kamke's ODE numbers for which {\bf symgen} found symmetries: 77\%.}\\
\end{tabular}}
\end{center}}
In summary, {\bf symgen} fails in $23 \%$, finds one symmetry in $37 \%$,
and finds more than one symmetry in $ 40 \%$ of the examples. As a
curiosity, although {\bf symgen}'s routines are also prepared to look for
symmetries not of polynomial or rational type, it is noticeable that most of
Kamke's 2\nd order non-linear examples have polynomial symmetries.

\subsection{Test of {\bf odsolve}'s integration schemes and comparison of
performances}
\label{comparison}

As explained in Sec.\ref{integrating}, for 2\nd order ODEs the knowledge
of one or more symmetries may or may not lead to the desired answer. For
some cases the routines may only obtain a reduction of order. So, to
evaluate the solving capabilities of the symmetry scheme here presented we
ran a performance test for {\bf odsolve} too, and compared the results
with those obtained by using the standard Maple {\bf dsolve} in solving
these 246 Kamke non-linear examples. The results we found can be
summarized as follows:

{\begin{center}
{\footnotesize
\begin{tabular}{|p{1.4 in}c|c|c|c|}
\hline
{\bf symgen} found & &
{\bf dsolve} solved & {\bf odsolve} solved & {\bf odsolve} reduced  \\
\hline
many symmetries for:    & 99 ODEs   & 57 &  98 &  1 \\
\hline 
1 symmetry for:         & 92 ODEs   & 18 &  36 & 52 \\
\hline 
0 symmetry for:         & 55 ODEs   &  0 &   0 &  0 \\
\hline 
Totals:                 & 246 ODEs  & 75 & 134 & 53 \\
\hline 
\multicolumn{5}{c}
{Table 3. \it Comparative performance}\\
\end{tabular} }
\end{center}}
Although Kamke's set of ODEs is just a particular one, some conclusions
can be drawn from the table above. First of all, the standard technique of
classifying the ODE according to whether it matches a given pattern (for
which we know the solving method) did not perform better in any example.
On the other hand, the implemented symmetry scheme, which tackles the ODEs
without making any distinction between them, almost doubled {\bf
dsolve}'s performance in obtaining closed form solutions, apart from being
able to return a reduction of order in more 53 examples. Also, {\bf
odsolve} succeeded in integrating to the end 36 examples for which only 1
symmetry was found, turning clear the relevance of implementing
integration alternatives for these cases. Finally, 98 of 99 examples for
which {\bf symgen} obtained more than 1 symmetry were solved by {\bf
odsolve} to the end, convincing us that the problem is now mainly related
to finding these symmetries, not their posterior use in the integration
process.

Concerning the cases in which {\bf symgen} found one or more pair of
infinitesimals but {\bf odsolve} fails in integrating or reducing the order
of the ODE, the problem was mainly related to: it was not possible to find
the inverse of the transformation from {\it canonical} variables $(r,s(r))$
to $(x,y(x))$, or this inverse involved {\bf RootOf}s and integrals. This
was the case in Kamke's ODEs 203, 204, 206, 238 and 246. As an example,
consider Kamke's ODE number 203,
\begin{equation}
\ode
a\,y\l(y-1\r){\frac {d^{2}y}{d{x}^{2}}}
-\l(a-1\r)\l(2\,y-1\r)\l({\frac { dy}{dx}}\r)^{2}
+f(x)\,y\l(y-1\r){\frac {dy}{dx}}y=0
\end{equation}
for which {\bf symgen} found one symmetry
\begin{verbatim}
> X := symgen(ode);
\end{verbatim}

\begin{equation}
{\rm X := }
\bigql  \xi
= 0,\ \ 
\eta  = 
\l({y}^{2}-y\r)
{\rm hypergeom}\l([1,{\frac {2}{a}}],[{\frac{a+1}{a}}],y\r)
+
\frac
{\l({y}^{{\frac{2\,a-1}{a}}}-{y}^{{\frac {a-1}{a}} }\r)}
{\l(1-y\r)^{{\frac{1}{a}}}}
\bigqr
\end{equation}
The corresponding canonical variables are given by:

\begin{verbatim} > canoni(X,y(x),s(r));
\end{verbatim}
\begin{eqnarray}
\bigcl r \r. & = & x,
\\
s(r) & = & \l. {\int^{y(x)}}\l(\left ({\it
\_z}^2-{\it \_z}\r) {\rm hypergeom}([1,{\frac {2}{a}}],[{\frac
{a+1}{a}}],{\it \_z}) +\frac {\l({{\it \_z}}^{{\frac {2\,a-1}{a}}}-{{\it
\_z}}^{{\frac {a-1}{a}}}\r)} {\l(1-{\it \_z}\r)^{\frac{1}{a}}}
\r)^{-1}{d\_z} \bigcr 
\nonumber
\end{eqnarray}
To reduce the order using these canonical variables the system solved them
w.r.t $\{x,y(x)\}$, arriving at expressions involving the {\bf RootOf} an
integral; even when the change of variables using those expressions was
successful, due to the complicated resulting algebraic structure, it was
impossible to verify the reduction of order, and the subroutine gave up with
no result.

\subsection{Test of symgen with the 1\st order Kamke examples}
\label{first_order}

Although the main purpose of this paper is to present the extension of the
package's commands to work with 2\nd order ODEs, some improvements have
been done in the abilities for solving 1\st order ODEs too. An update of
the results obtained by symgen for the 1\st order Kamke's examples is as
follows:

{\begin{center} {\footnotesize
\begin{tabular}
{|p{1cm}|c|c|c|c|c|c|c|c|c|c|c|}
\hline
 & & & \multicolumn{2}{|c|}{Average time} &
\multicolumn{7}{|c|}{$\xi$ and $\eta$ can be determined by} \\
\cline{4-12}
File & ODEs & Successes & {\it ok} & {\it fail} & \abaco{1} & 2 & 3 & 4 & 5
& 6 & \abaco{2} \\
\hline 
1 &  90 & 63 &  14.0 sec. &  7.7 sec. & 33 & 14 & 21 & 25 & 32 & 24 & 35 \\
\hline 
2 & 100 & 88 &  2.5 sec. &  1.6 sec. & 27 & 37 & 48 & 47 & 40 & 45 & 44 \\
\hline 
3 &  99 & 86 &  1.2 sec. & 8.2 sec. & 34 & 73 & 56 & 54 & 50 & 1 & 38 \\
\hline 
4 &  89 & 80 & 3.5 sec. & 16.3 sec. & 41 & 40 & 47 & 47 & 56 & 0 & 44 \\
\hline 
5 &  89 & 77 &  3.3 sec. & 15.7 sec. & 14 & 7  & 68 & 72 & 69 & 0 & 38 \\
\hline 
\hline 
Total: & 467 & 394 & $ \approx 5$ sec. & $\approx 10$ sec. & 149 & 171 &
240 & 245 & 247 & 70 & 199\\
\hline 
\multicolumn{12}{c}
{Table 4. \it Kamke's (non-quadrature) ODEs, infinitesimals determined by 
{\bf symgen}: $86 \%$}\\
\end{tabular} }
\end{center}}
The previous score (see Table 1. in \cite{odetools1}) was $75\%$, and the
improvement is mainly due to adjustments in {\bf `symgen/2`}, which is now
obtaining the infinitesimals for many rational ODEs, and to the introduction
of a new subroutine, {\bf `symgen/6`}, for tackling ODEs of Riccati type.
The number and class of Kamke's 1\st order ODEs for which {\bf symgen} fails
in determining a pair of infinitesimals are now given by:

{\begin{center} {\footnotesize
\begin{tabular}{|l|p{5.2in}|}
\hline
Class      &  Kamke's numbering (1\st order ODE examples)\\
\hline 
rational    &
%
203, 205, 234, 257, 432, 480, 482
   \\
\hline
d'Alembert     &
388, 390, 405, 406, 409, 410, 430, 479
   \\
\hline
Riccati     &
%
22, 25, 110
   \\
\hline
Abel        &
%
36, 37, 38, 40, 42, 43, 45, 46, 47, 48, 49, 50, 51, 111, 145, 146, 147, 
151, 169, 185, 219, 237, 250, 253, 265, 269
   \\
\hline
NONE        &
%
53, 54, 55, 56, 74, 79, 80, 81, 82, 83, 87, 121, 128, 197, 202, 206,
331, 340, 345, 351, 367, 370, 395, 456, 460, 461, 493
   \\
\hline 
\multicolumn{2}{c}{Table 5. \it
Kamke's 1\st order ODEs for which {\bf symgen} fails: $14 \%$}\\
\end{tabular}}
\end{center}}

\section{Computer algebra implementations of Lie symmetry methods}
\label{compare}

There are various computer algebra packages for performing Lie symmetry
analysis of differential equations \cite{hereman,www.can}. It is then
appropriate to situate ODEtools in the framework of the packages already
existing in Maple and other computer algebra systems (CAS)\footnote{The
information present in this section was gathered from
\cite{hereman,ibragimov,www.can,wolf,gonzalez1}}.

\subsection*{The range of applicability of ODEtools}

To start with, ODEtools is a package for tackling only first and second
order ODEs, while some of the other symmetry packages, as for example CRACK
and PDELIE \cite{www.can}, also handle systems of ODEs and PDEs of, in
principle, arbitrary order. The project is to extend ODEtools' commands to
handle high order ODEs soon, and concerning PDEs, ODEtools is presently
being developed as an ODE-package project, but it is fully integrated with
the PDEtools package \cite{pdetools}.

\subsection*{Dynamical symmetries and the solving scheme: input=ODE,
output=closed-form-solution}

Concerning first and second order ODEs, the goals in ODEtools are in some
sense a bit different from those of other symmetry packages. In ODEtools the
focus is mainly in returning closed form solutions to ODEs instead of in
performing a complete symmetry analysis. Thus, the idea implemented is to
look for {\it enough} symmetries of {\it point or dynamical} type as to be
able to integrate the ODE - instead of looking for {\it all} possible {\it
point} symmetries. Concretely, the difference is that most of the other
symmetry packages restrict the search to point symmetries\footnote{
Exception made by CRACK; restricted capabilities to determine dynamical
symmetries are also found in some other packages \cite{hereman}.}- and are
not prepared to build a closed form solution to the ODE from the knowledge
of these symmetries (see \cite{hereman,www.can}). It is worth mentioning
here that the search for point symmetries is more systematic due to the
existence of an overdetermined system of PDEs as departure point, but most
of second order ODEs don't have point symmetries. On the other hand all ODEs
have infinite symmetries of dynamical type. Also, though the determination
of the dimension of the symmetry group (done by other packages) before
searching for the symmetries may be useful, this is of no concrete help when
searching for dynamical symmetries since the dimension of the group is
infinite.


Related to these points, the ODEtools command for determining the symmetries
({\bf symgen}) does not restrict the type of the symmetries it looks for,
and the ODE-solver ({\bf odsolve}) has built-in routines for using both
types of symmetries to directly return a closed form solution for the
problem (see \se{description}), thus allowing the scheme {\it input=ODE,
output=answer} in a single step as shown in the examples throughout this
paper. Furthermore, the program is able to proceed with the integration
process of a second order ODE even when only one symmetry (of any type) was
found, by using it to reduce the ODE's order and restarting the solving
process with this reduced ODE as input. For the case in which the reduced
ODE cannot be solved, an appropriate scheme for returning reductions of
order was implemented. As an example of all this, consider

\begin{equation}
{\frac {d^{2}y}{d{x}^{2}}}
-{\frac{1}{y} {\left ({\frac {dy}{dx}}
\right )^{2}}}-\sin(x) \l({\frac {dy}{dx}}\r) y-\cos(x)\, y^{2}
\end{equation}

\ni This ODE does not have point symmetries \cite{gonzalez1}, and the
routines here presented only succeed in finding a single dynamical symmetry:

\begin{verbatim}
> symgen(ode,way=5);
\end{verbatim}
\begin{displaymath}
[\xi=0,\ \eta={\frac {\sin(x){y}^{2}\y1+\cos(x){y}^
{3}-{y}^{2}-{\y1}^{2}}{y}}]
\end{displaymath}

\ni The differential invariants corresponding to the infinitesimals above
are given by:

\begin{equation}
[I_0=x,I_1=\sin(x)y+{\frac {\y1}{y}}]
\end{equation}

\ni and their introduction as variables lead to the following first
integral

\begin{equation}
C_1= -{\frac{1} {y}\l({{\frac {dy}{dx}}+\sin(x)\,y^2}\r)}
\end{equation}

\ni with known symmetry (see \se{gon}). Finding the solution of the above
equation is then straightforward. In ODEtools, all these steps are run by
giving a single input:

\begin{verbatim}
> odsolve(ode);
\end{verbatim}
\begin{displaymath}
y={\frac {1+{C_1}^{2}}{-\cos(x)+\sin(x)C_1+{e^{-C_1\,x}}C_2+{e^{-
C_1\,x}}C_2\,{C_1}^{2}}}
\end{displaymath}



As a second example, let's consider the ODE shown in \cite{wolf}
illustrating the use of CRACK in solving ODEs using symmetry analysis:

\begin{equation}
\ode
3\,{\rho}^{2}h{\frac {d^{2}h}{d{\rho}^{2}}}-5\,{\rho}^{2
}\left ({\frac {dh}{d\rho}}\right )^{2}
+ \l(5\,\rho\,h-20\,\rho\,h^{3}\r) {\frac {dh}{d\rho}}
-20\,h^{4}+16\,h^{6}+4\,h^{2}=0
\label{example2}
\end{equation}

This ODE resulted from an attempt to generalize Weyl's class of solutions
of Einstein field equations \cite{kubitza}. According to \cite{wolf}, to
solve this ODE in the framework of CRACK and related set of packages
(LIEPDE, APPLYSYM, etc.) one should:

\begin{enumerate}

\item call LIEPDE to determine the ODE symmetries;

\item call APPLYSYM with the ODE and its symmetries and run an interactive
process to obtain a new ODE of reduced order;

\item use the second symmetry to integrate by hand the reduced ODE of the
previous step, to arrive at a parametric solution.\label{hand_made}

\end{enumerate}

\ni In the framework of ODEtools, all these steps are compacted into a
single call to {\bf odsolve}\footnote{ODEtools includes user-level commands
to run each one of these steps as well, see \se{package}.}. Also, many
integration strategies using the same symmetries in different manners are
optionally available. Using differential invariants for instance (see
\se{dif}), {\bf odsolve} directly arrives at an explicit solution for
\eq{example2} in terms of the roots of a quartic polynomial:

\begin{verbatim}
> odsolve(ode, dif);
\end{verbatim}
\begin{equation}
h=3\,{\frac {C_1\,{\rm RootOf}(3\,{{\it \_Z}}^{4}{\rho}^{2}-{
C_1}^{2}\left (1+2\,{\rho}^{2}C_2\right ){{\it \_Z}}^{2}+9\,{\rho}^{2}
)\rho}{{C_1}^{2}\left (1+2\,{\rho}^{2}C_2\right )\left ({\rm RootOf}(
3\,{{\it \_Z}}^{4}{\rho}^{2}-{C_1}^{2}\left (1+2\,{\rho}^{2}C_2\right 
){{\it \_Z}}^{2}+9\,{\rho}^{2})\right )^{2}-18\,{\rho}^{2}}}
\end{equation}

These roots can be made explicit as well, resulting in an explicit answer
without {\bf RootOf}. In this example, {\bf odsolve} spent 5.7 sec calling
{\bf symgen} to determine the symmetries, and 7 sec more to find the
differential invariants, change variables to reduce \eq{example2} to a first
order ODE with known symmetries, integrate the latter, and return to the
original variables. Moreover, in this case the two symmetries mentioned in
\cite{wolf} are already in normal form (i.e., they form a closed
subalgebra), but in other cases, the step (\ref{hand_made}) mentioned some
paragraphs above would not be possible in a direct manner since the reduced
ODE will not have the same symmetries as the original one. In such a case
{\bf odsolve} will automatically either rewrite the symmetries so that they
form a closed subalgebra, or determine a third symmetry when that is not
possible\footnote{In these cases, for second order ODEs, a third symmetry
can always be determined \cite{stephani}.}.

These differences do not mean that one package is better than the other -
and CRACK is one of the most powerful and complete symmetry packages
available - but emphasize that the goals and strategy of each of the
symmetry packages available are not completely equivalent. In CRACK or the
Macsyma package PDELIE, for instance, there are routines for handling
problems completely out of the scope of ODEtools as to search for the
symmetries of systems of differential equations.

\subsection*{The symmetries as particular solutions to the determining PDE
\eq{xi_eta}}

The computational strategy used in ODEtools for finding the symmetries is
also different from the strategy implemented in most of the other symmetry
packages. The difference is that the ODEtools routines directly look for
particular solutions for the determining PDE \eq{xi_eta}, instead of
building a set of overdetermined PDEs and trying to solve it exactly
\cite{odetools1}. Part of the motivation for this different approach is that
an overdetermined set of PDEs for the infinitesimals can only be built when
one restricts the search to point symmetries and this restriction is not
present in ODEtools. Another reason is that even restricting the search to
point symmetries, when a solution exists, this solution usually comes up
only after completely solving the whole system of PDEs. Of course this
requires the appropriate routines for ``solving systems of linear PDEs" but
this is a major problem on its own, and there is as yet no general algorithm
to integrate such a system \cite{hereman}. Moreover, most of the schemes for
tackling systems of PDEs will work only if the system of PDEs satisfies {\it
a priori} some restrictive conditions - e.g., the functional dependence on
the dependent variables should be of polynomial type or the routines only
work with rational function coefficients.

On the other hand, the search for particular solutions to \eq{xi_eta}
permits searching for point and dynamical symmetries in equal footing, by
reducing the problem to one of algebraic type or to auxiliary first and
second order ODEs as explained in our previous work \cite{odetools1}, and
without requiring routines for solving coupled systems of PDEs\footnote{For
the case of point symmetries, and provided that symmetries of
polynomial/rational type exist, there is an interesting work due to G. Reid
and McKinnon for finding particular solutions to the determining system of
PDEs by solving auxiliary ODEs \cite{reid1}.}. Moreover, there are {\it no
restrictions} (e.g., coefficient fields, functional dependence etc.) to the
type of ODE which can be tackled, and as a matter of fact, at least with
Kamke's non-linear examples, the average time consumed to solve an ODE is
very acceptable (see Sec.\ref{tests}), including in that both the
determination {\it and} the use of the symmetries for building a closed form
solution.


As an example of this, consider the determination of the symmetries for
Liouville's ODE

\begin{equation}
\ode {\frac {d^{2}y}{d{x}^{2}}}+g(y)\left ({\frac {dy}{dx}}\right)^{2}
+f(x){\frac {dy}{dx}}=0
\end{equation}

\ni where $f(x)$ and $g(y)$ are arbitrary functions - i.e., the ODE has
arbitrary dependence in both the independent and dependent variables. Such
a problem is out of the scope of CRACK (the ODE is not polynomial in the
dependent variable), and it is unsolvable using the reduce package SPDE,
which doesn't work with arbitrary functions \cite{hereman}. It takes $4.8\
\mbox{sec}$ to {\bf symgen}'s subroutines to determine the following two
particular solutions (in this case two point symmetries) to the
determining PDE \eq{xi_eta}:

\begin{verbatim}
> sym := symgen(ode);
\end{verbatim}
\begin{displaymath}
[\xi=0,\
\eta=\left (1+\int \!{e^{\int \!g(y){dy}}}{dy}\right ){e^{-\int
\!g(y){dy}}}],\ \ 
[\xi=\left (1-\int \!{e^{-\int \!f(x){dx}}}{dx}\right )
{e^{\int \!f(x){dx}}},\eta=0]
\end{displaymath}

and, in this case, the first of the symmetries above is enough to, in two
extra seconds, integrate the ODE constructively using canonical coordinates
(see \se{description}) via

\begin{verbatim}
> odsolve(ode, can);
\end{verbatim}

\begin{displaymath}
\ln \l(\int ^{y(x)}\!{e^{\int \!g({\it \_b}){d{\it \_b}}}}{d{\it \_b}}+1
\r)-\ln \l(\int \!{e^{-\int \!f(x){dx}}}{dx}+C_1\r)-C_2=0
\end{displaymath}

%


\subsection*{First order ODEs}

In the framework of other symmetry packages, Lie methods are usually
discarded in the case of first order ODEs. The reason is that the
determining PDE for the infinitesimals does not split into an overdetermined
system of PDEs, and most of the available symmetry packages are based on the
splitting of this PDE. On the other hand in the framework of ODEtools, the
approach presented in \cite{odetools1} put Lie methods as a concrete
alternative for tackling non-classifiable first order ODEs, and showed a
performance in solving Kamke's first order examples of $\approx 85 \%$ (see
\se{first_order}), representing the highest performance we are aware of in
solving these examples using symmetries or any other methods. Of course the
ODEtools scheme will fail with many other examples as well, but until a
general solution to the problem is discovered, this performance is
convincing us that the approach is worthwhile and a computer algebra
implementation of Lie methods can be a powerful alternative also for first
order ODEs.

\subsection*{Second order linear ODEs}

Another feature of the approach here presented is that it is possible to
tackle linear second order ODEs using the Lie scheme, while in other
symmetry packages these ODEs are excluded from the symmetry analysis process
(e.g., SPDE), or tackled in a manner which does not represent a concrete
advantage. The point is that when looking for point symmetries by trying to
solve an overdetermined system of PDEs, other symmetry packages would look
for symmetries of the form $[\xi=0, \eta=F(x)]$, where $F(x)$ is in turn a
particular solution to the homogeneous ODE. To find this type of symmetry is
then so complicate as solving the original ODE, and consequently many
authors tend to disregard the use of Lie methods for linear ODEs. On the
other hand, when looking for particular solutions to \eq{xi_eta}, these
symmetries $[\xi=0, \eta=F(x)]$ may arise naturally, as in \eq{linear_ODE},
and symmetries which {\it are not particular solutions} of the original ODE
can be found as well. For example, consider Kamke's second order linear ODE
number 142:

\begin{equation}
\ode 5\,\left (ax+b\right ){\frac {d^{2}y}{d{x}^{2}}}
+8\,a{\frac {dy}{dx}}+c\sqrt [5]{ax+b}\ y = 0
\end{equation}

\ni In this example,
a point symmetry which is not of the form $[\xi=0, \eta=F(x)]$ is given
by\footnote{This symmetry is determined by the internal subroutines as a
dynamical symmetry

$[0,F(x,y,\y1)]$
\begin{displaymath}
[\xi=0,\ \eta={\frac {\left
(5\,ax+5\,b\right )\y1+3\,ay}{5 \left (ax+b \right )^{3/5}a}}]
\end{displaymath}
\ni In the implementation here presented, the additional
conversion of dynamical symmetries into point symmetries, when possible,
is automatic.} :

\begin{verbatim}
> sym := symgen(ode);
\end{verbatim}
\begin{displaymath}
[\xi=-{\frac {\left (ax+b\right )^{2/5}}{a}},\ 
\eta={\frac {3 y}{5 \left (ax+b\right )^{3/5}}}]
\end{displaymath}

\ni Now all linear homogeneous ODEs have the symmetry $[\xi=0, \eta=y]$, so,
in the case of second order ODEs, only one more symmetry as in above is
enough to completely solve the problem. Although for linear ODEs the
standard schemes of rational, exponential, or special function solutions
seem to be the most appropriate methods, an implementation of Lie's method
by looking for particular solutions to \eq{xi_eta} appears as an interesting
complement since it can handle ODEs with arbitrary functions or solve the
problem when tricky changes of variables are involved. 

\subsection*{Self-contained environment for first and second order ODEs}

As the last difference mentioned in this summary, the ODEtools package has
the intention of being a self contained environment for solving and
studying first and second order ODEs. For this reason, the project
includes a compact on-line manual of solving methods for ODEs, implemented
here as the {\bf odeadvisor} command, and commands and options playing the
role of {\it tools} for tackling ODEs which cannot be solved using the
defaults.

Among the most relevant of such tools we can mention the {\bf dchange}
program, for performing general changes of variables in mathematical
expressions; the {\bf buildsym} program for determining the symmetries of an
ODE departing {\it from its solution}, and the {\bf equinv} program, for
determining the most general ODE simultaneously invariant under a given set
of point and/or dynamical symmetries. In a typical non-directly solvable
ODE, one would first try some changes of variables using {\bf dchange}, and
to determine the symmetries using the HINT option of {\bf symgen} as
explained in \se{description}\footnote{This possibility of participating in
the search for the symmetries when desired is a handy and useful feature,
almost absent in most of the other symmetry packages.}. When nothing works,
one can still consider a similar or simpler problem for which the solution
is known, try to build the corresponding symmetries using the {\bf buildsym}
command, and then try to understand what would be the symmetry pattern of
the unsolvable ODE. In connection with this, using the {\bf equinv} command
and departing from the symmetries returned by {\bf buildsym} one may be able
determine a family of ODEs containing the unsolvable ODE. Finally, if the
ODE is classifiable, the {\bf odeadvisor} command may also be of help,
giving either bibliographical references or some computational hints (e.g.,
possibly useful changes of variables etc.) to study the problem.

\section{Conclusions}

This paper presented a computer algebra implementation of symmetry methods
for solving 1\st and 2\nd order ODEs. This implementation proved to be a
valuable tool for tackling ODEs as shown in sec.\ref{tests}, resulting in
the extension to Maple's {\bf dsolve} we were looking for. Actually,
although the implementation of the method involves much more complicate
operations than the traditional matching-pattern schemes, the generality of
the families of ODEs which can be concretely solved with it is rather
impressive\footnote{The ODE-solver of ODEtools is now the core of the new
ODE-solver of the incomming Maple V Release 5.}.

Moreover, the solver can be used in an {\it interactive} manner, that is,
one can participate in the solving process (the {\tt HINT} and {\tt way=xxx}
options, the {\it gun} variables and the integration schemes; see
\se{description}), achieving in this manner a significant extension of the
solving capabilities of the scheme. This possible user participation plays a
fundamental role, because the functional form for the infinitesimals may be
out of the scope of {\bf symgen}'s defaults, and because {\bf odsolve}'s
defaults, even for the integration schemes (see \se{integrating}), may not
be the best choice for {\it all} cases.

Also, two routines, {\bf equinv} and {\bf buildsym}, were designed to tackle
the inverse problems, that is to find the most general ODE simultaneously
invariant under a given set of symmetries, and to find the symmetries of an
{\it unknown} ODE when its solution is given. These commands, together
with the HINT option of {\bf odsolve}, can be useful in many situations,
and for investigating new solving methods. Finally, the set of {\it
ODEtools} commands can be used with educational purposes too, since almost
all the relevant steps of the symmetry cycle are available as user-level
commands, and classification routines for most of the 1\st and 2\nd order
ODEs are available through the {\bf odeadvisor} command.

On the other hand, we also perceived that the complexity of the operations
involved in the symmetry scheme resulted in a slower solving method if
compared with the speed of the standard methods based on matching pattern
routines. This difference exists when solving 1\st order ODEs (but is almost
insignificant), becomes more noticeable when tackling 2\nd order ODEs, and
turns out to be relevant when considering high order ODEs. We then think
that an ODE-solver should use both methods, combined in an appropriate
manner in order to achieve the best performance. Concerning what would be
the best combination of methods, it is clear that a testing arena should be
defined first, and the results will depend on this choice. A collection of
testing examples most related to real problems seems to be the appropriate
choice, and in this sense Kamke's book appears to be one of the best
candidates available. Then, although the natural extension of this work
would be to implement symmetry methods for higher order ODEs, we think that
it is more convenient to invest in fine tuning the merging of symmetry and
standard methods for the 1\st/2\nd order cases first. This will open the way
for more efficient implementations for the high order case, as well as
permitting us to start more concrete work to tackle PDEs. We are
presently working in some prototypes in these directions\footnote{See
http://dft.if.uerj.br/odetools.html} and expect to succeed in obtaining 
reportable results in the near future.

\section*{Acknowledgments}

\noindent This work was supported by the State University of Rio de
Janeiro (UERJ), Brazil, and by the University of Waterloo, Ontario,
Canada. The authors would like to thank K. von B\"ulow\footnote{Symbolic
Computation Group of the Theoretical Physics Department at UERJ - Brazil.}
and T. Kolokolnikov\footnote{Symbolic Computation Group, Faculty of
Mathematics, University of Waterloo - Canada.} for a careful reading of
this paper, A.D. Roche\footnotemark[10] for preparing a first version of
help pages for 2\nd order ODEs; and Prof. M.A.H. MacCallum\footnote{Queen
Mary and Westfield College, University of London - U.K.} for kindly
sending us valuable references.


\vskip -2cm
\begin{thebibliography}{99}

\bibitem{kamke} Kamke, E., {\it Differentialgleichungen: 
L{\"o}sungsmethoden und L{\"o}sungen}. Chelsea Publishing Co, New York 
(1959).

\bibitem{odetools1} E.S. Cheb-Terrab, L.G.S. Duarte and L.A.C.P. da Mota,
``Computer Algebra Solving of First Order ODEs Using Symmetry Methods",
Computer Physics Communications, 101 (1997) 254.

\bibitem{hereman} W. Hereman, Chapter 13 in vol 3 of the CRC Handbook of
Lie Group Analysis of Differential Equations, Ed.: N.H.Ibragimov, CRC
Press, Boca Raton, Florida (1995).

\bibitem{gon} A. Gonz\'ales L\'opez, {\it Phys.Lett.} {\bf A133} (1988),
190.

\bibitem{olver2} P.J. Olver, {\it Equivalence, Invariants, and Symmetry},
Cambridge University Press (1995).

\bibitem{zwillinger} D. Zwillinger, {\it Handbook of Differential
Equations}, 2\nd edition, Academic Press (1992).

\bibitem{stephani} H. Stephani, {\it Differential equations: their
solution using symmetries}, ed.\ M.A.H. MacCallum, Cambridge University
Press, New York and London (1989). 

\bibitem{bluman} G.W. Bluman and S. Kumei, {\it Symmetries and Differential
Equations}, Applied Mathematical Sciences {\bf 81}, Springer-Verlag (1989).

\bibitem{olver} P.J. Olver, {\it Applications of Lie Groups to Differential
Equations}, Springer-Verlag (1986).

\bibitem{pdetools} E.S. Cheb-Terrab and K. von B\"ulow, Comp. Phys. Comm. 
90 (1995), 102-116. For the last version of this package see
http://dft.if.uerj.br/pdetools.html.

\bibitem{ibragimov} CRC ``Handbook of Lie Group Analysis of Differential
Equations", V.1: {\it Symmetries, Exact Solutions and Conservation Laws},
ed. N.H. Ibragimov, CRC Press, Boca Raton (1994).

\bibitem{www.can} Links to web pages related to most of the computer
algebra packages for symmetry analysis are found at:
http://www.can.nl/Systems\_and\_Packages/Per\_Purpose/Special.


\bibitem{wolf} T. Wolf, {\it Examples of the investigation of differential
equations with modularized programs},
ftp://euclid.maths.qmw.ac.uk/pub/crack/demo.ps

\bibitem{gonzalez1} A. Gonz\'alez-L\'opez, {\it Symmetry and integrability 
by quadratures of ordinary differential equations}, Phis. Lett. A, (1988) 
190.

\bibitem{kubitza} M. Kubitza, FSU Jena, private communication.

\bibitem{reid1} {\sc G.J. Reid} and {\sc D.K. McKinnon},\, {\em Solving
systems of linear PDEs in their coefficient field by recursively decoupling
and solving ODEs}.\, Preprint, Department of Mathematics\, (University of
British Columbia, Vancouver, Canada, 1993). Submitted to J. Symb. Comp.
(1993)


\end{thebibliography}
\end{document}